%% file: main.tex
\documentclass{SciPost}

\hypersetup{
    colorlinks,
    linkcolor={red!50!black},
    citecolor={blue!50!black},
    urlcolor={blue!80!black}
}

\usepackage[colorinlistoftodos,prependcaption,textsize=small]{todonotes}

\newcommand{\SU}{\ensuremath{\mathrm{SU}}}

\newcommand{\Ugroup}{\ensuremath{\mathrm{U}}}  

\let\oldepsilon\epsilon       
\let\oldvarepsilon\varepsilon 
\renewcommand{\epsilon}{\oldvarepsilon}
\renewcommand{\varepsilon}{\oldepsilon}
\newcommand{\dblquote}[1]{``\text{#1}''}

\usepackage[bitstream-charter]{mathdesign}
\DeclareSymbolFont{usualmathcal}{OMS}{cmsy}{m}{n}
\DeclareSymbolFontAlphabet{\mathcal}{usualmathcal}

\fancypagestyle{SPstyle}{
\fancyhf{}
\lhead{\colorbox{scipostblue}{\bf \color{white} ~SciPost Physics }}
\rhead{{\bf \color{scipostdeepblue} ~Submission }}

\fancyfoot[C]{\textbf{\thepage}}
}

\input{Packages_and_Functions}

\begin{document}

\pagestyle{SPstyle}

\begin{center}{\Large \textbf{\color{scipostdeepblue}{
Exploring the freeze-in mechanism for scotogenic singlet dark matter\\
}}}\end{center}

\begin{center}\textbf{
Ugo de Noyers\textsuperscript{1$\star$} and 
Bj\"orn Herrmann\textsuperscript{1$\dagger$} 
}\end{center}

\begin{center}
{\bf 1} LAPTh, Univ.\ Savoie Mont Blanc, CNRS, F-74000 Annecy, France
\\[\baselineskip]
$\star$ \href{mailto:denoyers@lapth.cnrs.fr}{\small denoyers@lapth.cnrs.fr} \qquad
$\dagger$ \href{mailto:herrmann@lapth.cnrs.fr}{\small herrmann@lapth.cnrs.fr}
\end{center}

\section*{\color{scipostdeepblue}{Abstract}}
\textbf{\boldmath{%
We present an exploratory study of the freeze-in mechanism within a scotogenic framework, where dark matter can either be a scalar singlet or a fermion singlet. Based on a random parameter scan, we show that large portions of the parameter space feature a dark matter relic density in agreement with the limits derived by Planck. Moreover, constraints related to lepton flavour violation are mostly respected within such parameter regions. Our study shows that it will be worth to further investigate the freeze-in mechanism within scotogenic frameworks.
}}

\vspace{\baselineskip}

\noindent\textcolor{white!90!black}{%
\fbox{\parbox{0.975\linewidth}{%
\textcolor{white!40!black}{\begin{tabular}{lr}%
  \begin{minipage}{0.6\textwidth}%
    {\small Copyright attribution to authors. \newline
    This work is a submission to SciPost Physics. \newline
    License information to appear upon publication. \newline
    Publication information to appear upon publication.}
  \end{minipage} & \begin{minipage}{0.4\textwidth}
    {\small Received Date \newline Accepted Date \newline Published Date}%
  \end{minipage}
\end{tabular}}
}}
}


\vspace{10pt}
\noindent\rule{\textwidth}{1pt}
\tableofcontents
\noindent\rule{\textwidth}{1pt}
\vspace{10pt}

\input{./tex_files/intro.tex}
\input{./tex_files/models.tex}
\input{./tex_files/setup.tex}
\input{./tex_files/results.tex}
\input{./tex_files/conclusion.tex}

\section*{Acknowledgements}
The authors would like to thank G.~Bélanger, A.~Goudelis, and W.~Porod for helpful discussions, particularly concerning the use of {\tt SARAH}, {\tt SPheno} and {\tt micrOMEGAs}. B.\,H.\ would like to thank R.\ Mandal for interesting discussions in a very early stage of this work. The plots presented in this article have been obtained using {\tt MatPlotLib} \cite{MatPlotLib}.

\paragraph{Funding information}
The work of U.\,d.\,N.\ is funded by Ph.D.\ grant of the French Ministry for Education and Research. The authors acknowledge funding from the French \emph{Programme d’investissements d’avenir} through the Fédération Enigmass+.


\bibliography{main}

\end{document}

%% file: Packages_and_Functions.tex
\usepackage{bm}                             
\usepackage{slashed}                        
\usepackage{physics}                        
\usepackage{mathtools}                      

\usepackage{epstopdf}   
\usepackage{subcaption} 
\usepackage{float}      

\usepackage{array}
\usepackage{booktabs}    
\usepackage{multirow}
\usepackage{tabularx}    

\usepackage{pifont}      

\usepackage{tikz}        
\usetikzlibrary{positioning}            
\usetikzlibrary{shapes.geometric}       
\usetikzlibrary{arrows.meta}
\usetikzlibrary{decorations.pathmorphing} 

\usepackage{rotating}    
\usepackage{pgfplots}    
\pgfplotsset{compat=1.18} 
\usepackage{pgf-pie}     

\usepackage[compat=1.1.0]{tikz-feynman}
\tikzfeynmanset{warn luatex=false}

\let\oldepsilon\epsilon       
\let\oldvarepsilon\varepsilon 

\renewcommand{\epsilon}{\oldvarepsilon}
\renewcommand{\varepsilon}{\oldepsilon}

\usepackage{tikz,pgfplots}

\usetikzlibrary{positioning} 
\usetikzlibrary{arrows.meta} 
\usetikzlibrary{calc} 

\colorlet{myred}{red!80!black}
\colorlet{myblue}{blue!80!black}
\colorlet{mygreen}{green!50!black}
\colorlet{myorange}{orange!80!yellow!90!red!90!black}

\tikzstyle{mycomment}=[inner sep=1pt,scale=0.75,align=left]

\tikzstyle{mybox}=[draw,#1!80!black,fill=#1!95!black!20,inner sep=5pt,outer sep=3pt,
                   thick,rounded corners=3pt,align=center,font=\bfseries]

\tikzstyle{myarrow}=[-{Latex[length=8,width=8]},#1!80!black,thick,line cap=round,line width=3]

\def\connect[#1](#2)!#3!(#4){
  \draw[#1] (#2) |- ($(#2)!#3!(#4)$) node[pos=0.5] (#2-#4-1) {}
  -| (#4) node[pos=0.5] (#2-#4-2) {}
}

%% file: tex_files/intro.tex
\section{Introduction}
\label{Sec:Intro}

Although the Standard Model (SM) of particle physics has proven to be a highly successful theory in many aspects, several open questions are left unanswered. In addition to questions of theoretical nature, there are two issues related to observations, which cannot be explained within the SM alone. First, the SM does not include a suitable candidate particle for the cold dark matter (CDM), which is observed to account for about 25\% of the Universe's energy content \cite{Planck2018}. Second, the neutrinos are strictly massless in the SM, while the observed neutrino oscillations \cite{NuFit2024} imply that at least two mass eigenvalues are non zero. In order to address (at least some of) the above shortcomings, the SM needs to be extended by additional fields, including at least one viable dark matter candidate, and providing a mechanism for neutrino mass generation.

Concerning DM, most viable candidates are stabilized by introducing an additional discrete $\mathbb{Z}_2$ symmetry, under which the SM fields are even, while additional fields are odd. In such a framework, today's DM relic abundance can be explained through two distinct mechanisms, namely the \emph{freeze-out} \cite{KolbTurnerEarlyUniverse, GondoloGelmini1990} or the \emph{freeze-in} \cite{Hall:2009bx} mechanism. In the former, DM is produced in thermal equilibrium in the early Universe as \dblquote{Weakly Interacting Massive Particles} (WIMPs). Their initially relatively high abundance then decreases during the Universe's evolution due to the expansion of the Universe and DM (co-)annihilation into SM particles. 

In the freeze-in picture, the DM particle is never in thermal equilibrium and is produced through scattering or decays of SM particles during the evolution of the Universe. Such a mechanism requires rather small couplings between the SM bath and the DM particle, the latter being names \dblquote{Feebly Interacting Massive Particle} (FIMP) in this case. In contrast to the freeze-out case, the FIMP production rate can also depends on the reheating temperature \cite{Hall:2009bx, Bringmann:2021sth}.

Regardless of the exact mechanism, to be a phenomenologically viable framework, the underlying model has to satisfy the constraint stemming from recent observations of today's CDM relic density,
\begin{align}
	\Omega_{\rm CDM}h^2 ~=~ 0.1206 ~\pm~ 0.0012 \,,
	\label{Eq:Planck}
\end{align} 
given in the $\Lambda$CDM cosmological model at the 68\% confidence level \cite{Planck2018}, or at least not overshoot the upper limit. While the freeze-out mechanism has been studied over the last decades within many classes of beyond the Standard Model frameworks, the freeze-in picture has emerged more recently \cite{Hall:2009bx}.

Coming to neutrino mass generation, let us first recall that the neutrinos, in contrast to the other SM fermions, are not subject to the Higgs mechanism, as there are no right-handed neutrinos in the SM. Moreover, the neutrino masses are known to be relatively small, their sum being constrained by cosmological considerations to \cite{DESI2024}
\begin{align}
	\sum_{i=1,2,3} m_{\nu_1} ~\leq~ 0.072 ~{\rm eV}\,. 
	\label{Eq:NuMasses1}
\end{align}
at the 95\% confidence level. While the exact mass eigenvalues are experimentally inaccessible, assuming normal ordering, at the 68\% confidence level, the squared mass differences are constraint to \cite{NuFit2024}
\begin{align}
	\begin{split}
	\Delta m^2_{12} ~=~ m_{\nu_2}^2 - m_{\nu_1}^2 ~=~ \big( 7.49 \pm 0.19 \big)\cdot 10^{-5} ~{\rm eV}^2 \,, \\
	\Delta m^2_{13} ~=~ m_{\nu_3}^2 - m_{\nu_1}^2 ~=~ \big( 2.51 \pm 0.02 \big)\cdot 10^{-3} ~{\rm eV}^2 \,.
	\end{split}
	\label{Eq:NuMasses2}
\end{align}
Note that, although this possibility may seem somewhat unnatural, experimental data leaves the possibility of having one massless neutrino, i.e.\ $m_{\nu_1} = 0$.

In many studies, the necessary mass terms are introduced via the so-called Seesaw mechanism, where typically heavy right-handed neutrinos are postulated as additional particles \cite{Minkowski1977}. An alternative is to consider the generation of neutrino masses through one-loop diagrams including suitable scalar and fermionic fields. Let us note that, in this case, despite the loop suppression of the mass terms, the couplings between the lepton doublets and the new fields are required to be reasonably small in order to explain the smallness of the resulting neutrino masses. Let us also note that, independently of the chosen mechanism for mass generation, lepton flavour violating interactions are introduced as the Pontecorvo-Maki-Nakagawa-Sakata (PMNS) matrix is not diagonal any more. Such processes are experimentally constrained, especially in the case of $\mu-e$ transitions, where stringent limits arise from recent experimental data \cite{PDG2024} with the perspective of even lower limits in a near future \cite{Meucci:2022qbh, Moritsu:2022lem}.
 
In the present work, we consider the possibility of the freeze-in mechanism within so-called ``scotogenic'' extensions, originally proposed in Ref.\ \cite{Ma2006}, where additional scalar and fermion fields are added, which are odd under a discrete $\mathbb{Z}_2$ symmetry. While the lightest electrically neutral among the new eigenstates is a viable dark matter (DM) candidate, neutrino masses are generated via one-loop diagrams involving the new fields. More specifically, we consider a specific scotogenic framework, including a scalar singlet and a fermionic singlet, which are supposed to be the possible DM candidates. In this way, gauge interactions of the DM particle are avoided, and the relevant couplings can be kept numerically small enough to favour the out-of-equilibrium situation and thus DM production via the freeze-in mechanism. 

The present work aims at exploring the possibility of realizing the observed DM relic density within such a framework. Moreover, we will show that lepton-flavour violating processes, which are naturally present with scotogenic frameworks, are naturally suppressed in the case of freeze-in DM production. In this rather exploratory than extensive study, we employ a simple random parameter scan to convieniently cover the available parameter space. More sophisticated exploration methods, such as a Markov Chain Monte Carlo scan (as in, e.g., Refs.\ \cite{Sarazin:2021nwo, Alvarez:2023dzz, deNoyers:2024qjz}, or a Deep Neural Network (as in, e.g., Ref.\ \cite{deNoyers:2025ije}) are left for future work. Previous studies of scotogenic FIMP DM remain scarce \cite{Molinaro:2014lfa, Ma:2022bfa, Berbig:2022nre, Shibuya:2022xkj}, but encouraging to explore this possibility.

The present paper is organized as follows: We introduce the chosen scotogenic model in Sec.\ \ref{Sec:Model}, detailing particular the Lagrangian, the associated parameters, and the resulting physical mass spectrum. In the following Sec.\ \ref{Sec:Constraints} we present the computational setup of our study, together with the observables we are interested in. The results of our study are presented and discussed in Sec.\ \ref{Sec:Results}. Conclusions are given in Sec.\ \ref{Sec:Conclusion}.

%% file: tex_files/models.tex
\section{The model Lagrangian and mass spectrum}
\label{Sec:Model}

We consider a scotogenic framework, where the SM gauge group is extended by a $\mathbb{Z}_2$ symmetry, under which all SM fields are even, while additional fields are odd. The field content is extended by a real scalar singlet $S$, a complex scalar doublet $\eta$, a fermion singlet $F$ and two generations of the same fermion triplet $\Sigma_1$ and $\Sigma_2$. These extra fields do not carry colour charge, i.e.\ they are $\SU(3)_{\text{C}}$ singlets. A summary of the respective representations under $\SU(2)_{\text{L}} \times \Ugroup(1)_{\text{Y}}$ is given in Table \ref{Tab:T3QuantumNumbers}. This specific scotogenic framework belongs to the \dblquote{T3} category according to the classification given in Ref.\ \cite{Yaguna2013}. It shares the scalar sector with the \dblquote{T1-2A} and \dblquote{T1-2G} models, while its fermionic sector is the combination of the fermion sectors of the \dblquote{T3-B} and \dblquote{T3-C} models.

\begin{table}
    \centering
    \begin{tabular}{|c||c|c||c|c|}
    \hline
            \     & ~$F$~ & ~$\Sigma_k$~ & ~$S$~ & ~$\eta$~ \\
            \hline
            \hline
        $\SU(2)_{\mathrm{L}}$   &  $\mathbf{1}$ & $\mathbf{3}$ & $\mathbf{1}$ & $\mathbf{2}$ \\
        \hline
        $\Ugroup(1)_{\mathrm{Y}}$  & 0 & 0 & 0 & 1 \\
        \hline
    \end{tabular}
    \caption{The additional fields of the scotogenic model under consideration in the present work. The index $k$ runs over the two fermion triplet generations, $k=1,2$.}
    \label{Tab:T3QuantumNumbers}
\end{table}

The physical mass eigenstates of the theory are obtained by diagonalizing the mass matrices arising from the Lagrangians detailed below. In practice, we use the numerical spectrum calculator {\tt SPheno~4.0.5} \cite{SPheno2003, SPheno2011} to obtain the physical mass spectrum including corrections at the one-loop level. The model has been implemented using the {\tt Mathematica} package {\tt SARAH~4.15.1} \cite{SARAH2008, SARAH2010, SARAH2011, SARAH2013, SARAH2014}.

\subsection{Scalar sector: Lagrangian and parameters}
\label{Subsec:Scalars}

The scalar sector of the model under consideration consists of the SM Higgs doublet $H$, and two additional scalar fields: the singlet $S$ and the doublet $\eta$. Upon electroweak symmetry breaking (EWSB), only $H$ acquires a vacuum expectation value $v = \sqrt{2} \langle H \rangle \approx 246$ GeV. In component notation, the two doublets are then written as
\begin{equation}
    H ~=~ \begin{pmatrix} G^+ \\ \frac{1}{\sqrt{2}} \big( v + h^0 + i G^0 \big) \end{pmatrix}, \qquad
    \eta ~=~ \begin{pmatrix} \eta^+ \\ \frac{1}{\sqrt{2}} \big( \eta^0 + i A^0 \big) \end{pmatrix} \,.
    \label{Eq:ScalarDoublets}
\end{equation}
Here, $G^0$ and $G^+$ are the Goldstone bosons and $h^0$ the physical Higgs boson. The new doublet $\eta$ contains a charged scalar $\eta^+$, a neutral $CP$-even scalar $\eta^0$, and a neutral $CP$-odd pseudoscalar $A^0$. Note that, as we suppose the $\mathbb{Z}_2$ symmetry to be unbroken, the doublet $\eta$ does not acquire a vacuum expectation value. The scalar sector is the same as for the scotogenic \dblquote{T1-2A} \cite{Sarazin:2021nwo, Alvarez:2023dzz, Darricau:2025vcs} and \dblquote{T1-2G} \cite{deNoyers:2024qjz, deNoyers:2025ije} frameworks.

The Lagrangian of the scalar sector of this model is given by 
\begin{equation}
    \begin{split}
        - \mathcal{L}_{\text{scalar}} ~&=~ M^2_H \vert H \vert^2 + \lambda_H \vert H \vert^4 
        + \frac{1}{2} M^2_S S^2 + \frac{1}{2} \lambda_{4S} S^4 + M^2_{\eta} \vert \eta \vert^2 + \lambda_{4\eta} \vert \eta \vert^4 \\ 
        &~~~~+~ \frac{1}{2} \lambda_S S^2 \vert H \vert^2  + \frac{1}{2} \lambda_{S \eta} S^2 \vert \eta \vert^2 + \lambda_{\eta} \vert \eta \vert^2 \vert H \vert^2 +  \lambda^{\prime}_{\eta} \vert \eta^{\dagger} H \vert^2 \\ 
        &~~~~+~ \frac{1}{2} \lambda^{\prime\prime}_{\eta} \Big( \big( \eta^{\dagger} H  \big)^2 + \text{h.c.} \Big) + \kappa \left( S \eta^{\dagger} H + \text{h.c.} \right) \,,
    \end{split}
    \label{Eq:ScalarLagrangian}
\end{equation}
where we suppose all scalar couplings to be real. The first two terms of the second line are the SM terms related to the Higgs doublet $H$. At tree level, after EWSB, the usual minimization relation in the Higgs sector, 
\begin{equation}
    m^2_{h^0} ~=~ -2 M_H^2 ~=~ 2 \lambda_H v^2 \,,
    \label{Eq:HiggsMass}
\end{equation}
allows to eliminate the free mass parameter $M^2_H$ in favour of the Higgs self-coupling $\lambda_H$. Imposing $m_{h^0} \approx 125 \; \text{GeV}$ leads to a tree-level value of $\lambda_H \approx 0.13$.

The remaining terms of the second line are the mass terms and self-couplings of the additional singlet and doublet. The terms in the third and last line of Eq.\ \eqref{Eq:ScalarLagrangian} include all possible couplings between the Higgs doublet, the additional doublet and the additional singlet. Mixing between the singlet and the doublet is induced by the trilinear coupling $\kappa$.

\subsection{Scalar sector: Masses and mixing matrices}
\label{Subsec:ScalarMasses}

After EWSB, the neutral scalar states $S$ and $\eta^0$ mix into two $CP$-even neutral mass eigenstates, denoted $\phi_1^0$ and $\phi_2^0$. As we suppose the scalar couplings to be real, there is no mixing with the third neutral component $A^0$, which is $CP$-odd. In the basis $\{ S, \eta^0, A^0 \}$, the tree-level mass matrix is given by
\begin{equation}
    \mathcal{M}^2_{\phi} ~=~ \begin{pmatrix}
        M^2_S + \frac{1}{2} v^2 \lambda_S & v \kappa & 0 \\
        v \kappa & M^2_{\eta} + \frac{1}{2} v^2 \lambda_L & 0 \\
        0 & 0 & M^2_{\eta} + \frac{1}{2} v^2 \lambda_A
    \end{pmatrix} \,,
    \label{Eq:ScalarMatrix}
\end{equation}
where $\lambda_{L,A} = \lambda_{\eta} + \lambda^{\prime}_{\eta} \pm \lambda^{\prime \prime}_{\eta}$. The rotation to the physical mass basis is defined as 
\begin{equation}
    \left( \phi_1^0, \phi_2^0, A^0 \right)^t ~=~ U_{\phi} \, \left( S, \eta^0, A^0 \right)^t \,,
    \label{Eq:ScalarMixing}
\end{equation}
where the $CP$-even mass eigenstates are supposed to be ordered such that $m_{\phi^0_1} \leq m_{\phi^0_2}$. Finally, the tree-level mass of the charged scalar eigenstates ($\phi^{\pm}=\eta^{\pm}$) is given by
\begin{equation}
    m^2_{\phi^{\pm}} ~=~ M^2_{\eta} + \frac{1}{2} v^2 \lambda_{\eta} \,.
    \label{Eq:ChargedScalarMassEigenstates}
\end{equation}

\subsection{Fermion sector: Lagrangian and parameters}
\label{Subsec:Fermions}

In addition to the SM fermions, the scotogenic \dblquote{T3} framework considered in this study features a fermion singlet $F$ and two generations $\Sigma_{1,2}$ of a Majorana triplet. The latter can be represented as 
\begin{equation}
    \Sigma_k ~=~ \begin{pmatrix}
        \Sigma^0_k/\sqrt{2} & ~\Sigma^+_k \\ ~\Sigma^-_k & -\Sigma^0_k/\sqrt{2} 
    \end{pmatrix} \qquad \text{with}~ k \in \{1,2\} \,.
    \label{Eq:FermionTriplets}
\end{equation}
The corresponding fermionic Lagrangian,
\begin{equation}
    -\mathcal{L}_{\text{fermion}} ~=~ \frac{1}{2} M_{F} \bar{F} F  + \frac{1}{2} \sum_{i,j} M_{\Sigma_{ij}} \text{Tr}\big\{ \overline{\Sigma}_i \Sigma_j \big\} + \text{h.c.} \,,
    \label{Eq:FermionLagrangian}
\end{equation}
includes mass terms for the singlet, denoted $M_F$, and for the triplets $M_{\Sigma_{ij}}$ with $(i,j) \in \{1,2\}$. Note that there is no Yukawa term inducing mixing between the singlet and the triplets. Moreover, we place ourselves in a basis where $M_{\Sigma_{12}} = M_{\Sigma_{21}} = 0$, without a loss of generality, implying the absence of mixing between the two triplets.

\subsection{Fermion masses and mixing}
\label{Subsec:FermionMasses}

The mass spectrum of the fermion sector is given by the mass matrices
\begin{equation}
    M_{\chi^0} = \begin{pmatrix}
        M_{F} & 0 & 0 \\
        0 & M_{\Sigma_{11}} & 0\\
        0 & 0 & M_{\Sigma_{22}}
    \end{pmatrix} \, , \quad\quad
    M_{\chi^{\pm}} = \begin{pmatrix}
        M_{\Sigma_{11}} & 0 \\
        0 & M_{\Sigma_{22}}
    \end{pmatrix}
    \, ,
    \label{Eq:Fermion_Mass_Matrix}
\end{equation}
which are diagonal matrices due to the absence of mixing terms. The mass eigenstates of the fermion sector are equivalent to the interactions eigenstates, namely $\left(F, \Sigma_1^0, \Sigma_2^0, \Sigma_1^{+}, \Sigma_2^{+}\right)$. The ordering between the charged states will depend only on the hierarchy of the parameters $M_{\Sigma_{11}}$ and $M_{\Sigma_{22}}$. In view of our phenomenological study of freeze-in dark matter, the neutral states are ordered such that we ensure the lightest neutral state $\chi^0_1$ being the singlet $F$. The specific ordering for $\chi_2^0$ and $\chi_3^0$ follow the same pattern as the one discussed for the charged states.

\subsection{Interaction terms}
\label{Subsec:Interactions}

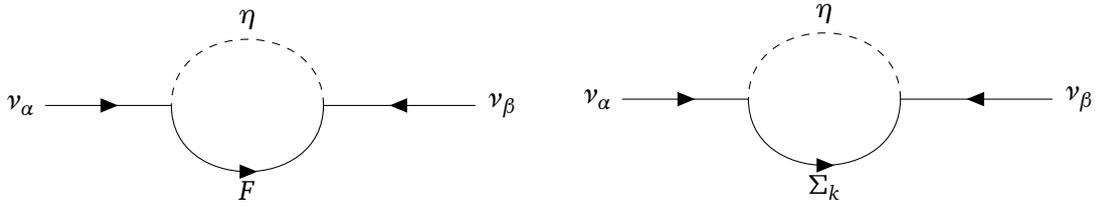
\begin{figure}[tb]
    \centering
    \begin{subfigure}{0.45\textwidth}
        \centering
        \begin{tikzpicture}
            \begin{feynman}
                \vertex (start) {$\nu_{\alpha}$};
                \vertex[right=2cm of start] (loopstart);
                \vertex[right=2cm of loopstart] (loopend);
                \vertex[right=2cm of loopend] (end) {$\nu_{\beta}$};
                \diagram* {
                    (start) -- [fermion] (loopstart),
                    (loopstart) -- [fermion, half right, edge label'=$F$] (loopend),
                    (loopend) -- [scalar,  half right, edge label'=$\eta$] (loopstart),
                    (end)   -- [fermion] (loopend),
                };
            \end{feynman}
        \end{tikzpicture}
    \end{subfigure}
    \qquad
    \begin{subfigure}{0.45\textwidth}
        \centering
        \begin{tikzpicture}
            \begin{feynman}
                \vertex (start) {$\nu_{\alpha}$};
                \vertex[right=2cm of start] (loopstart);
                \vertex[right=2cm of loopstart] (loopend);
                \vertex[right=2cm of loopend] (end) {$\nu_{\beta}$};
                \diagram* {
                    (start) -- [fermion] (loopstart),
                    (loopstart) -- [fermion, half right, edge label'=$\Sigma_k$] (loopend),
                    (loopend) -- [scalar,  half right, edge label'=$\eta$] (loopstart),
                    (end)   -- [fermion] (loopend),
                };
            \end{feynman}
        \end{tikzpicture}
    \end{subfigure}~~~
    \caption{Radiative generation of neutrino masses within the \dblquote{T3} model, depicted in the interaction basis. The indices $\alpha, \beta$ run over the neutrino flavours, $(\alpha, \beta) \in \{e, \mu, \tau\}$, while the index $k$ runs over the two fermionic triplet generations, $k \in \{1,2\}$.}
    \label{Fig:T3_Neutrino_Masses}
\end{figure}

The final part of the Lagrangian contains the interactions between the SM leptons and the additional fields introduced above,
\begin{equation}
    - \mathcal{L}_{\mathrm{int}} ~\supset~ g_F^{\alpha} F \eta L_{\alpha} ~+~ g_{\Sigma}^{\alpha j} \eta \Sigma_j L_{\alpha} ~+~ \mathrm{h.c.} \,.
    \label{Eq:InteractionLagrangian}
\end{equation}
Here, the lepton doublets are denoted $L_{\alpha}$ with $\alpha=e,\mu,\tau$, and sums over $\alpha$ and $j=1,2$ are implicit. Three non-zero neutrino masses are radiatively generated through the one-loop diagrams depicted in Fig.\ \ref{Fig:T3_Neutrino_Masses} corresponding to the terms in Eq.\ \eqref{Eq:InteractionLagrangian}. 

%% file: tex_files/setup.tex
\section{Constraints and computational setup}
\label{Sec:Constraints}

The addition of the fields and interactions presented in the previous section yields including a total of 27 additional parameters. In practice, to study the parameter space in view of the freeze-in dark matter phenomenology, we will randomly scan over these parameters. Let us recall that the viable dark matter candidates within the freeze-in picture are the lightest scalar or the lightest fermion, provided that they are singlet-like. In the fermion sector, this is naturally the case if $M_F < M_{\Sigma_{1,2}}$. In the scalar case, in addition to $M^2_S < M^2_{\eta}$, we need to prevent mixing by keeping the trilinear coupling $\kappa$ numerically small.

In addition to having singlet dark matter, the freeze-in mechanism requires to restrict the couplings between the dark matter candidate and the bath particles to values lower than ${\cal O}(10^{-10})$. Let us mention again that it is this actual requirement that excludes gauge interactions, and consequently closes the possibility of having doublet or triplet dark matter candidates in the freeze-in paradigm. 

For the following study, we randomly scan the parameter space according to the two sets of parameter ranges given in Table \ref{Tab:DM_parameter_range}, which match the aforementioned conditions and thus favour either scalar or fermion freeze-in dark matter. For scalar FIMPS, we namely impose $M_S < 2 M_{\eta}$ while keeping the relevant couplings numerically small, $|\lambda_S| \leq 10^{-10}$, $|\lambda_{S\eta}| \leq 10^{-10}$, and $|\kappa| \leq 10^{-10}$ GeV. Similarly, in the fermion case, we impose $M_F < M_{\Sigma_{1,2}}$ together with $|g_F^{\alpha}| \lesssim 10^{-10}$ ($\alpha=e,\mu,\tau$). In each case, the dark matter mass is considered to be in the interval between 200 GeV and 20 TeV. Note that we scan the SM Higgs coupling $\lambda_H$ around its tree-level value due to the impact of the included one-loop corrections. In order to gain efficiency in probing the maximum of the parameter space, we chose to scan all parameters on a logarithmic scale, and a possible sign is assigned on a random basis. Only the parameter sets featuring a normal ordering for neutrino masses will be considered.

\begin{table}
    \centering
    \begin{tabular}{c@{\hspace{0.5cm}}c@{\hspace{0.5cm}}c}
        \begin{tabular}{|c|c|}
            \hline
            \multicolumn{2}{|c|}{\bf Scalar dark matter} \\
            \hline
            \hline
            $\lambda_H$ & $[0.1, 0.4]$ \\
            \hline\hline
            $M_{S}$ & $[200, 20\,000]$ \\
            \hline
            $M_{\eta}$ & $[2, 3] \cdot M_S$ \\
            \hline
            $\lambda_{S}, \lambda_{S \eta}$ & $[-10^{-10}, 10^{-10}]$ \\
            \hline
            $\lambda_{4S}, \lambda_{4\eta}$ & $[10^{-9}, 1]$ \\
            \hline
            $\lambda_{\eta}, \lambda^{\prime}_{\eta}, \lambda^{\prime\prime}_{\eta}$ & $[-1, 1]$ \\
            \hline
            $\kappa$ & $[-10^{-10}, 10^{-10}]$ \\
            \hline
            \hline
            $M_{F}$ & $[2, 3] \cdot M_S$ \\
            \hline
            $M_{\Sigma_{jj}}$ & $[2, 3] \cdot M_S$ \\
            \hline
            $\mathfrak{Re}\{g^{\alpha}_{F}\}$ & $[-1, 1]$ \\
            \hline
            $\mathfrak{Im}\{g^{\alpha}_{F}\}$ & $[-1, 1]$ \\
            \hline
            $\mathfrak{Re}\{g^{\alpha j}_{\Sigma}\}$ & $[-1, 1]$ \\
            \hline
            $\mathfrak{Im}\{g^{\alpha j}_{\Sigma}\}$ & $[-1, 1]$ \\
            \hline        
            \hline
            $T_R$ & $[10^5, 10^8]$ \\
            \hline
        \end{tabular}
        & \quad & 
        \begin{tabular}{|c|c|}
            \hline
            \multicolumn{2}{|c|}{\bf Fermion dark matter} \\
            \hline
            \hline
            $\lambda_H$ & $[0.1, 0.4]$ \\
            \hline\hline
            $M_S$ & $[2, 3] \cdot M_F$ \\
            \hline
            $M_{\eta}$ & $[2, 3] \cdot M_F$ \\
            \hline
            $\lambda_{S}, \lambda_{S \eta}$ & $[-1, 1]$ \\
            \hline
            $\lambda_{4S}, \lambda_{4\eta}$ & $[10^{-9}, 1]$ \\
            \hline
            $\lambda_{\eta}, \lambda^{\prime}_{\eta}, \lambda^{\prime\prime}_{\eta}$ & $[-1, 1]$ \\
            \hline
            $\kappa$ & $[-1000, 1000]$ \\
            \hline
            \hline
            $M_{F}$ & $[200, 20\,000]$ \\
            \hline
            $M_{\Sigma_{jj}}$ & $[2, 3] \cdot M_F$ \\
            \hline 
            $\mathfrak{Re}\{g^{\alpha}_{F}\}$ & $[-10^{-10}, 10^{-10}]$ \\
            \hline
            $\mathfrak{Im}\{g^{\alpha}_{F}\}$ & $[-10^{-10}, 10^{-10}]$ \\
            \hline
            $\mathfrak{Re}\{g^{\alpha j}_{\Sigma}\}$ & $[-1, 1]$ \\
            \hline
            $\mathfrak{Im}\{g^{\alpha j}_{\Sigma}\}$ & $[-1, 1]$ \\
            \hline
            \hline
            $T_R$ & $[10^5, 10^8]$ \\
            \hline
        \end{tabular}
    \end{tabular}
    \caption{Parameters and associated ranges for a numerical random scan. In the case of scalar dark matter (left) the ranges are chosen so as to obtain a singlet scalar dark matter candidate, in the case of fermion dark matter (right) the chosen ranges lead to a singlet fermion dark matter candidate. All dimensionful parameter are given in GeV. The indices $\alpha$ and $j$ run over $\alpha=e,\mu,\tau$ and $j=1,2$, respectively.}
    \label{Tab:DM_parameter_range}
\end{table}

As already mentioned before, starting from a set of parameters chosen in the above intervals, we make use the of the numerical spectrum calculation {\tt SPheno} to obtain the physical mass spectrum including one-loop corrections. In case of an unphysical mass spectrum, e.g.\ in the case of tachyonic states, the parameter point is rejected. Moreover, we require the existence of a stable electroweak vacuum, i.e.\ the scalar potential needs to be bounded from below. This translates into a set of inequality conditions for the scalar couplings, which are in the present case identical to those presented in Refs.\ \cite{Sarazin:2021nwo, deNoyers:2024qjz} for the scotogenic \dblquote{T1-2A} and \dblquote{T1-2G} frameworks. Parameter sets failing these conditions are rejected as well. 

Having computed the physical mass spectrum for a given parameter point, we make use of the public dark matter code {\tt micrOMEGAs 6.1.15} \cite{MO2001, MO2004, MO2007a, MO2007b, MO2013, MO2018, MO2024} to compute the dark matter relic density $\Omega_{\rm CDM}h^2$. As for {\tt SPheno}, the implementation of the model has been realised through {\tt SARAH}, which provides model files for {\tt CalcHEP} \cite{CalcHep_Belyaev:2012qa}.  As discussed above, the reheating temperature is an ingredient to the FIMP production cross-section, that is not related to the particle content of the model. We impose $T_R \gtrsim 10^5$ GeV, so as to set it at least one order of magnitude above the other energy scales of the model. In our study, we find that the exact value of $T_R$ does not significantly impact the predicted relic density. 

In addition to the predicted value of $\Omega_{\rm CDM}h^2$, the code also returns the list of processes contributing to the associated production cross-section. Let us point out that we do not strictly require the relic density to fall within the ranges of Eq.\ \eqref{Eq:Planck}, but we rather aim at investigating to which extent a scotogenic framework can give raise to a viable freeze-in mechanism, and in addition study the dependence on the different key parameters.

The couplings $g_{F}^{\alpha}$ and $g_{\Sigma}^{\alpha j}$ ($\alpha \in \{e,\mu,\tau\}$ and $j \in \{1,2\}$), introduced in Eq.\ \eqref{Eq:InteractionLagrangian}, give rise to lepton-flavour violating interactions, mediated at the one-loop level. As currently, no direct evidence of such processes has been observed, the couplings are required to remain relatively small. 

Let us point out that the most stringent constraints stem from the search for $\mu - e$ transitions, such as the branching ratio of the decay $\mu \to e\gamma$, measured by the MEG experiment \cite{MEG:2016leq}, the decay $\mu \to 3e$, measured by Mu3e \cite{Blondel:2013ia}, and the $\mu-e$ conversion rate in nuclei, e.g.\ in Au, measured by SINDRUM II  \cite{SINDRUMII:2006dvw}, Mu2e, COMET, and DeeME \cite{Rule:2024kjo, Calibbi:2017uvl, Natori:2014yba}. Important improvements on the experimental sensitivity of certain observables will arrive in a very near future, namely with the MEG II \cite{Meucci:2022qbh} and the COMET ($\mu-e$ conversion in Al) \cite{Haxton:2022piv, Moritsu:2022lem, deGouvea:2013zba, Artikov:2023vsa} experiments. In our numerical scan, the observables related to lepton flavour violation as well as the electric dipole moments of leptons are computed using {\tt SPheno}.

Let us finally mention that, in our numerical study, the relevant SM parameters are fixed to $G_F = 1.166370 \cdot 10^{-5}$ GeV, $\alpha_S(M_Z) = 0.1187$, $M_Z = 91.1887$ GeV, $m_b(m_b) = 4.18$ GeV, $m_t^{\text{pole}} = 173.5$ GeV, and $m_{\tau}(m_{\tau}) = 1.77669$ GeV.

%% file: tex_files/results.tex
\section{Results and discussion}
\label{Sec:Results}

In this section, we present the results obtained from the random scan described above. While our main focus will lie on the dark matter relic density, we will also discuss the predictions on the level of lepton flavour violating processes. Both aspects give rise to discussions of the different coupling parameters. In the following, we will first discuss the case of scalar singlet dark matter, then the case of fermion singlet dark matter.

\subsection{Scalar Feebly Interacting Massive Particle}
\label{Subsec:Scalar_DM}

\begin{figure}[tb]
    \centering
    \includegraphics[width=0.48\textwidth]{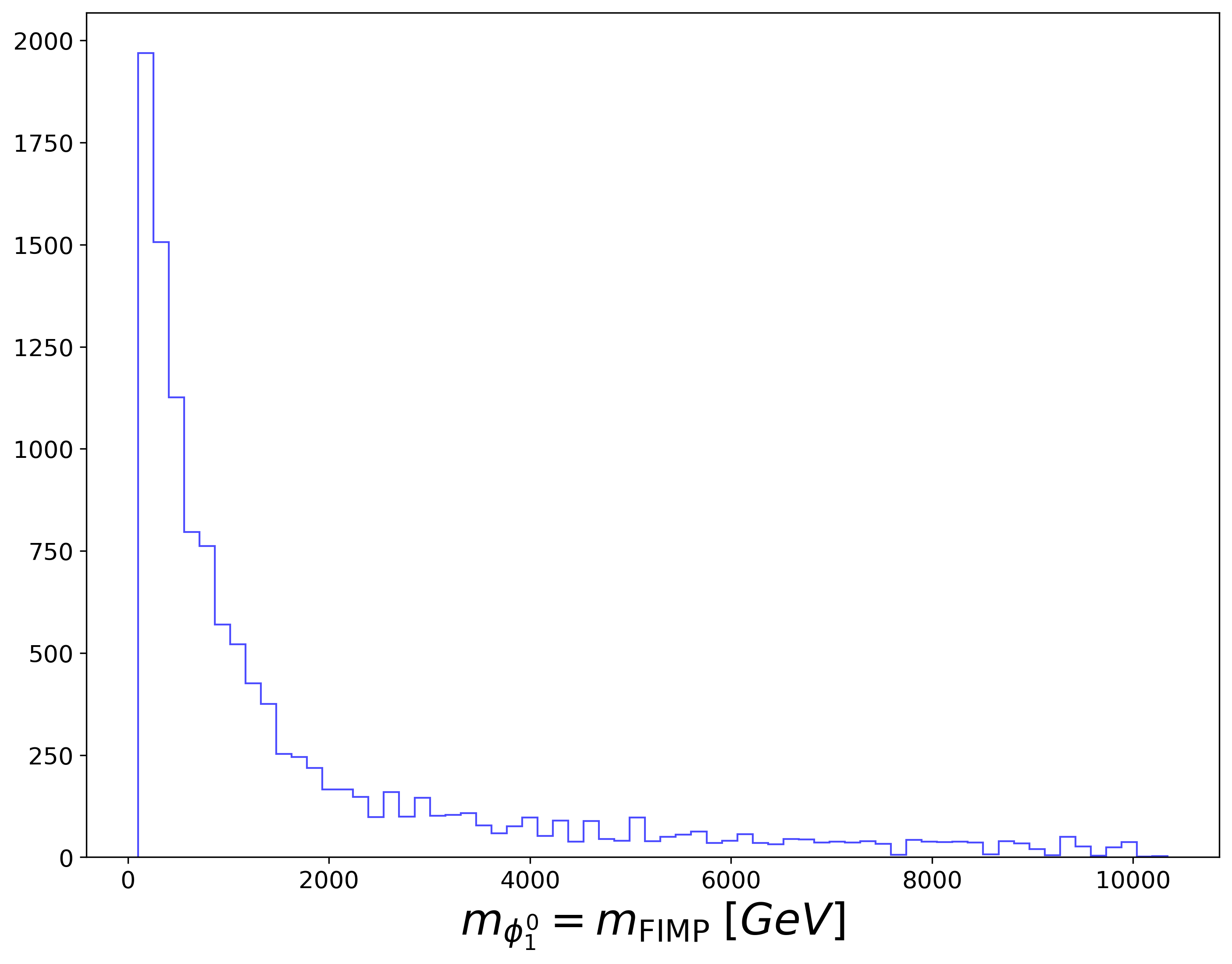}
    ~~~
    \includegraphics[width=0.47\textwidth]{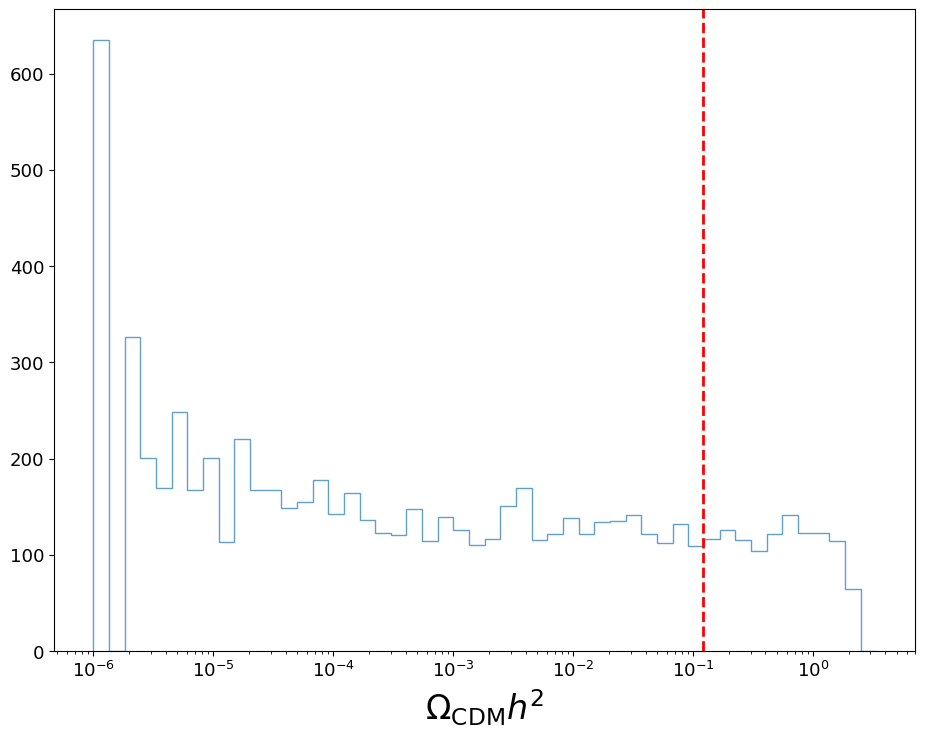}
    \caption{Mass distribution of the scalar FIMP $\phi_1^0$ (left) and distribution of the associated relic density $\Omega_{\rm CDM}h^2$ in the considered model obtained from a random scan yielding 11\,942 parameter points within the parameter space defined in Table \ref{Tab:DM_parameter_range}. The red line indicates the central value $\Omega_{\rm CDM}h^2 = 0.12$ of Eq.\ \eqref{Eq:Planck}.}
    \label{Fig:Scalar_DM:Hist_DM_mass_omh2}
\end{figure}

For the case of scalar singlet dark matter, our random scan has given a total of 11\,942 accepted parameter points. We start by showing the associated distributions of the dark matter mass as well as the relic density in Fig.\ \ref{Fig:Scalar_DM:Hist_DM_mass_omh2}. The mass distribution peakes at lower values of the allowed range, i.e.\ towards masses of around 200 GeV. This effect is triggered by the requirement that the singlet must be the lightest state, combined with the fact that the associated mass parameter has been scanned on a logarithmic scale. Our study covers mainly scalar dark matter masses between 200 GeV and 2000 GeV.

The relic density features a rather flat distribution, except for very low values, indicating an intrinsic preference for smaller production cross-sections within the framework under consideration. As can be seen, a large majority (about 90\%) of the accepted points are below the upper limit of about $\Omega_{\rm CDM}h^2 \lesssim 0.12$, implying that corresponding parameter configurations are valid in the sense that dark matter does not overclose the Universe. For parameter points yielding a relic density of about $\Omega_{\rm CDM}h^2$, the production cross-section is dominated by FIMP pair production from $W^+W^-$ pairs (about 50\%), from $Z^0 Z^0$ pairs (about 25\%), and from $h^0 h^0$ pairs (about 25\%). For certain point, minor contributions stem from $t\bar{t}$ initial states. Note that possible contributions to $\Omega_{\rm CDM}h^2$ from decays are absent in the present scenarios due to a too high FIMP mass.

Let us note that our study does not show any apparent correlation between the dark matter mass and the relic density. Moreover, we did not find any correlation with the reheating temperature not the ratio of the dark matter mass and the reheating temperature. This is coherent with the findings of Ref.\ \cite{Bringmann:2021sth} that the relic density is not sensitive to the reheating temperature if the latter becomes large enough with respect to the other quantities in the model, e.g.\ the FIMP mass.

\begin{figure}[tb]
    \centering
    \includegraphics[width=0.45\textwidth]{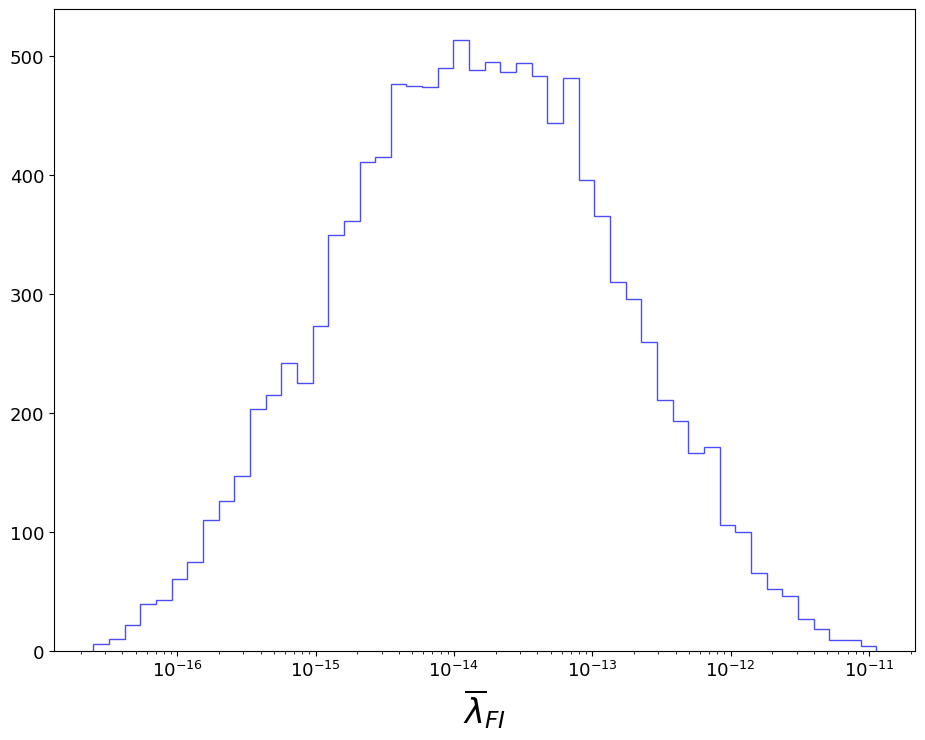}
    ~~~
    \includegraphics[width=0.475\textwidth]{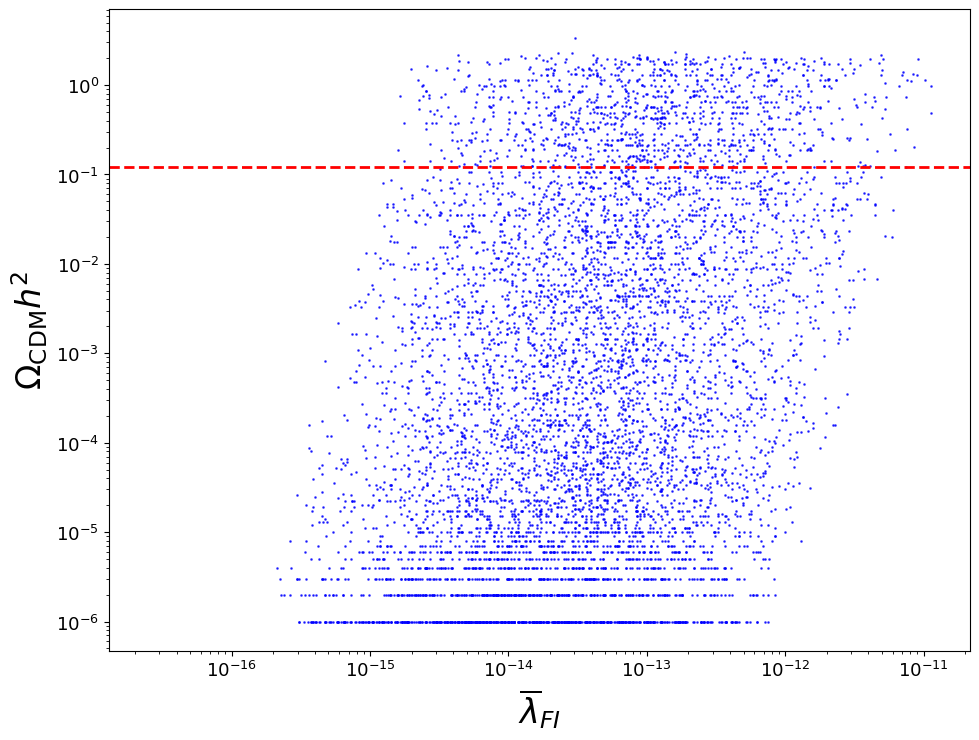}
    \caption{Left: Distribution of the geometric mean $\bar{\lambda}_{FI}$ of the couplings participating in the scalar FIMP production. Right: Correlation of the geometric mean of the scalar couplings $\bar{\lambda}_{FI}$ and the relic density $\Omega_{\rm CDM} h^2$ for the scalar FIMP $\phi^0_1$. Both plots are obtained from a random scan yielding 11\,942 viable parameter points featuring scalar FIMP DM.}
    \label{Fig:Scalar_DM:Scatter_geometrical_mean_lambda_omegah2}
\end{figure}

Coming to the couplings involved in the freeze-in production of dark matter, we examine the distribution of the geometric mean value 
\begin{equation}
    \bar{\lambda}_{FI} ~=~ \left(\lambda_S \lambda_{S\eta} \frac{\kappa}{v} \right)^{1/3} \; ,
    \label{Eq:Geometric_mean_lambda_FI}
\end{equation}
of the couplings of the scalar singlet FIMP, i.e.\ participating to the freeze-in production of the latter. The distribution obtained from our random scan is shown in the left panel of Fig.\ \ref{Fig:Scalar_DM:Scatter_geometrical_mean_lambda_omegah2}, while the right panel shows its correlation with the relic density. The mean value defined in Eq.\ \eqref{Eq:Geometric_mean_lambda_FI} peaks around $\bar\lambda_{\rm FI} \sim 10^{-13}$, which is identified as the typical couplings strength needed for successful freeze-in mechanism. Moreover, we observe a slight correlation with the predicted relic density, which is expected as the latter is proportional to the involved couplings \cite{Hall:2009bx}.

Let us finally discuss the phenomenology of lepton flavour violation in the framework under consideration. As already discussed above, the most stringent constraints stem from $\mu-e$ transitions, while transitions involving the $\tau$ lepton are typically less constraint. We focus here on the decays $\mu \to e\gamma$, $\mu \to 3e$, $\tau\to\mu\gamma$, and $\tau\to 3\mu$. In the present case, the branching ratios of these decays are governed by the couplings $g^{\alpha}_{F}$ and $g^{\alpha j}_{\Sigma}$ ($\alpha=e,\mu,\tau$, $j=1,2$). As already pointed out in Refs.\ \cite{Sarazin:2021nwo, deNoyers:2024qjz}, it is interesting to examine the geometrical mean value of the involved couplings, 
\begin{equation}
    \bar{\mathcal{G}} ~=~ \big| g_{F}^{e} g_{F}^{\mu} g_{F}^{\tau} g_{\Sigma_1}^{e} g_{\Sigma_1}^{\mu} g_{\Sigma_1}^{\tau} g_{\Sigma_2}^{e} g_{\Sigma_2}^{\mu} g_{\Sigma_2}^{\tau} \big|^{1/9} \, .
    \label{Eq:Modulus_geometric_mean_g}
\end{equation}

\begin{figure}[tb]
    \centering
    \includegraphics[width=0.45\textwidth]{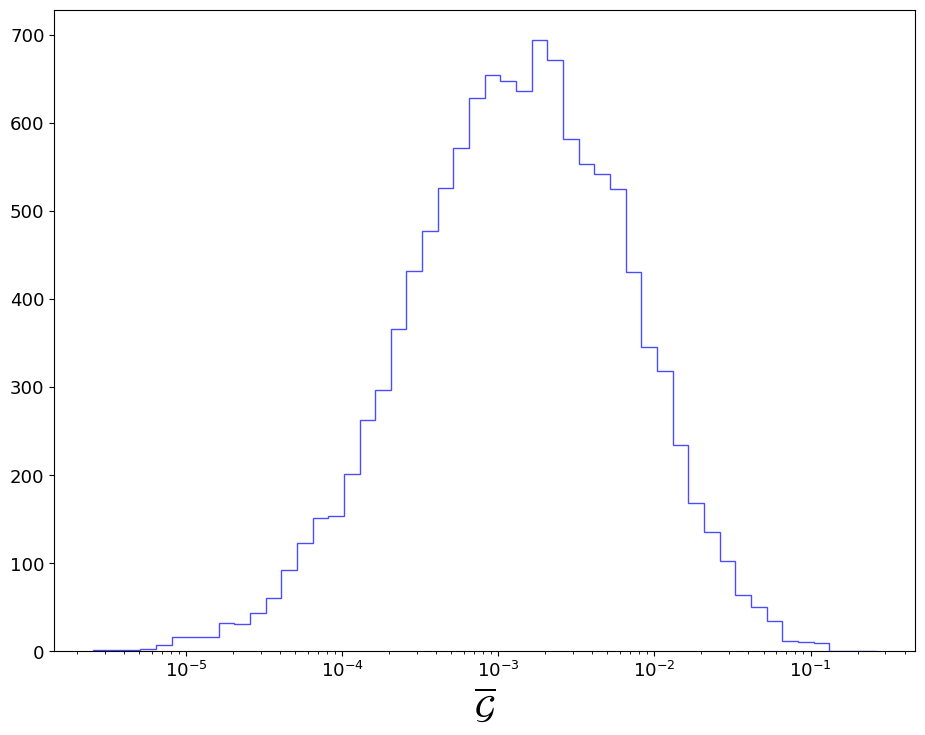}~~~~
    ~~~~
    \includegraphics[width=0.48\textwidth]{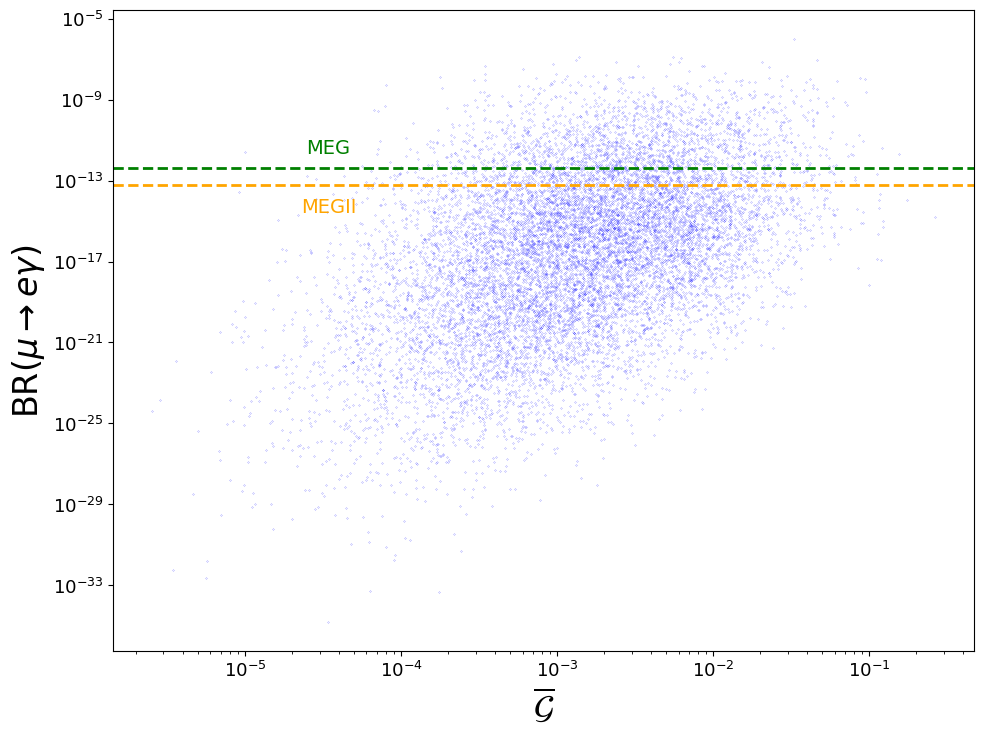}
    \caption{Left: Distribution of the modulus of the geometric mean of the couplings $\bar{\mathcal{G}}$ defined in Eq.\ \eqref{Eq:Modulus_geometric_mean_g}. Right: Correlation of the modulus of the geometric mean of the couplings $\bar{\mathcal{G}}$ and the branching ratio of the decay $\mu\to e\gamma$. Both plots are obtained from a random scan yielding 11\,942 viable parameter points featuring scalar FIMP DM.}
    \label{Fig:Scalar_DM:Histo_modulus_g_geometric}
\end{figure}    

\begin{figure}[h!]
    \centering
    \includegraphics[width=0.47\textwidth]{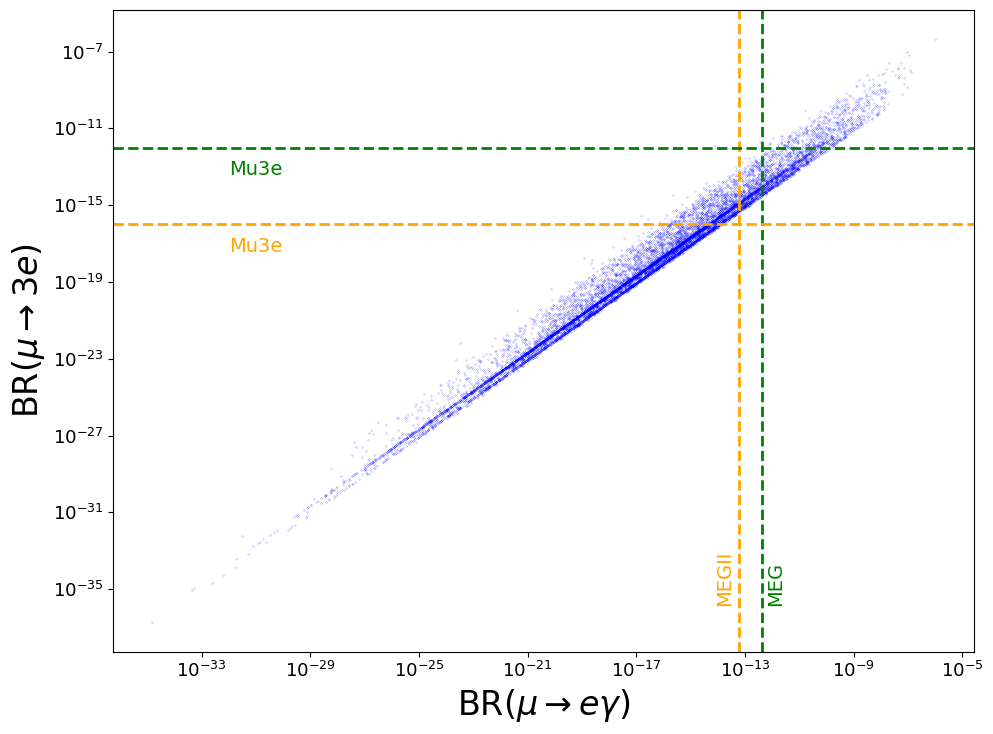}
    ~~~~\includegraphics[width=0.47\textwidth]{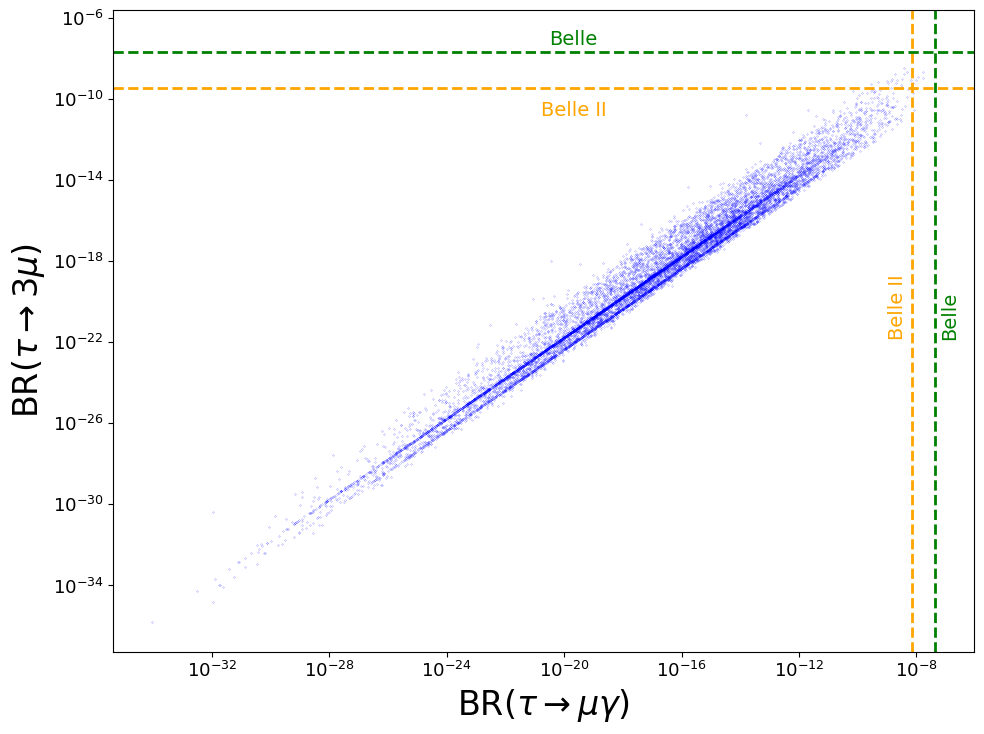}\\[2mm]
    \includegraphics[width=0.47\textwidth]{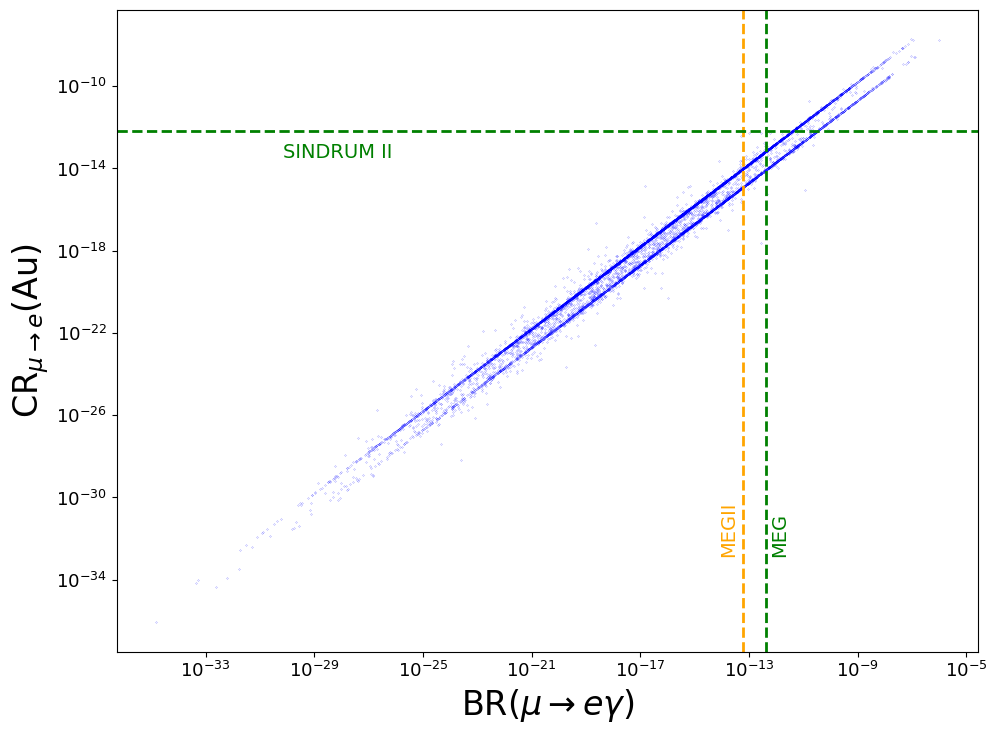}
    ~~~~\includegraphics[width=0.47\textwidth]{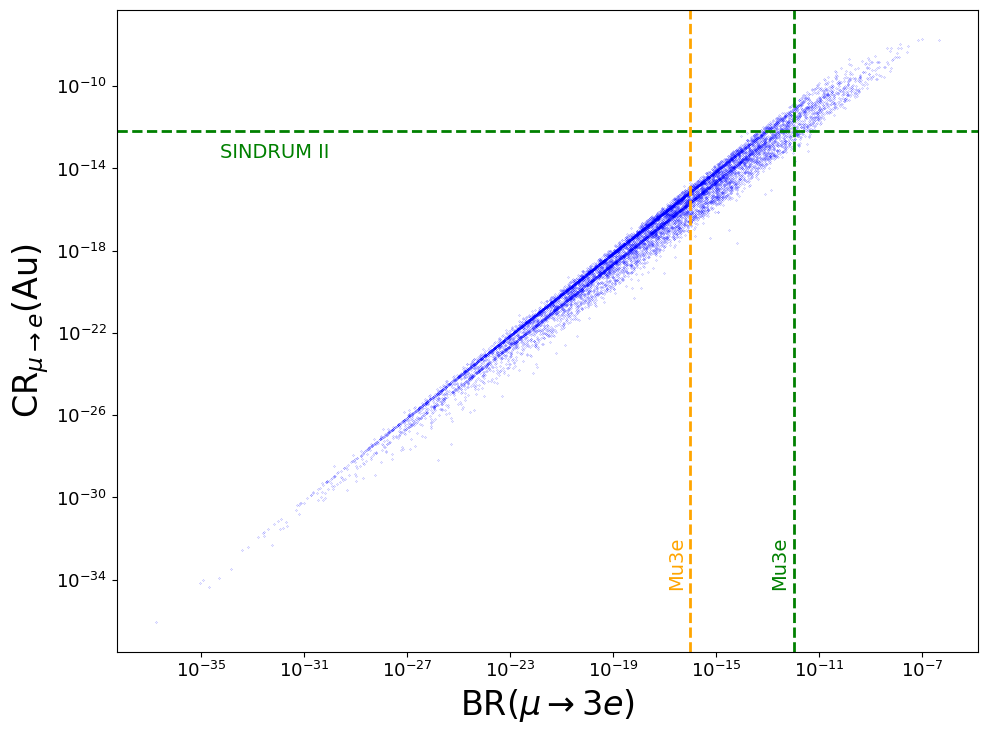}
    \caption{Upper left: Correlation of the respective branching ratios of the decays $\mu \to e \gamma$ and $\mu \to 3e$. Upper right: Same for the decays $\tau \to \mu \gamma$ and $\tau \to 3 \mu$. Lower left: Correlation between the $\mu-e$ CR in Au nuclei and BR of the decays $\mu \to e \gamma$. Lower right: Correlation between the $\mu-e$ CR in Au nuclei and BR of the decays $\mu \to  3e$. All plots are obtained from a random scan yielding 11\,942 viable parameter points featuring scalar FIMP DM. Green and orange dashed lines indicate current and future experimental limits, respectively.}
    \label{Fig:Scalar_DM:Scatter_BR_BR}
\end{figure}    

We show in the left part of Fig.\ \ref{Fig:Scalar_DM:Histo_modulus_g_geometric} the associated distribution, which peaks around $\bar{\cal G} \sim 10^{-3}$. Note that this mean value is numerically larger compared to the corresponding mean value obtained in Refs.\ \cite{Sarazin:2021nwo, deNoyers:2024qjz} for the \dblquote{T1-2A} and \dblquote{T1-2G} frameworks, where the peak is found rather about one order of magnitude lower around $10^{-4}$. In the right part of Fig.\ \ref{Fig:Scalar_DM:Histo_modulus_g_geometric}, we show the obtained values of $\bar{\cal G}$ against the corresponding values of  BR($\mu\to e\gamma)$. The graph displays a mild correlation, which is to some extent washed out by the influence of the other parameters. 

Despite this overall increase of the couplings, the lepton flavour violating observables are satisfied for a large number of parameter points (about 85\%), as can be seen in Fig.\ \ref{Fig:Scalar_DM:Scatter_BR_BR}, where we display several correlations between representative LFV observables. Moreover, we observe rather clean correlations, implying that triangle contributions are dominating the associated amplitudes. Note that in the freeze-out scenarios discussed in Refs.\ \cite{Sarazin:2021nwo, deNoyers:2024qjz}, the box contributions were found to be more important, washing out the correlations, in particular for the part of parameter space featuring higher values of the LFV branching ratios or conversion rates. Finally, note that a certain part of the parameter space (about 6\% of the remaining parameter sets) will be challenged by experiments starting in a relatively near future, such as MEG II \cite{Meucci:2022qbh}, COMET \cite{Artikov:2023vsa}, Mu3e \cite{Blondel:2013ia}, and Belle II \cite{Belle-II:2024sce}.

\subsection{Fermion Feebly Interacting Massive Particle}
\label{Subsec:Fermion_DM}

\begin{figure}[tb]
    \centering
    \includegraphics[width=0.47\textwidth]{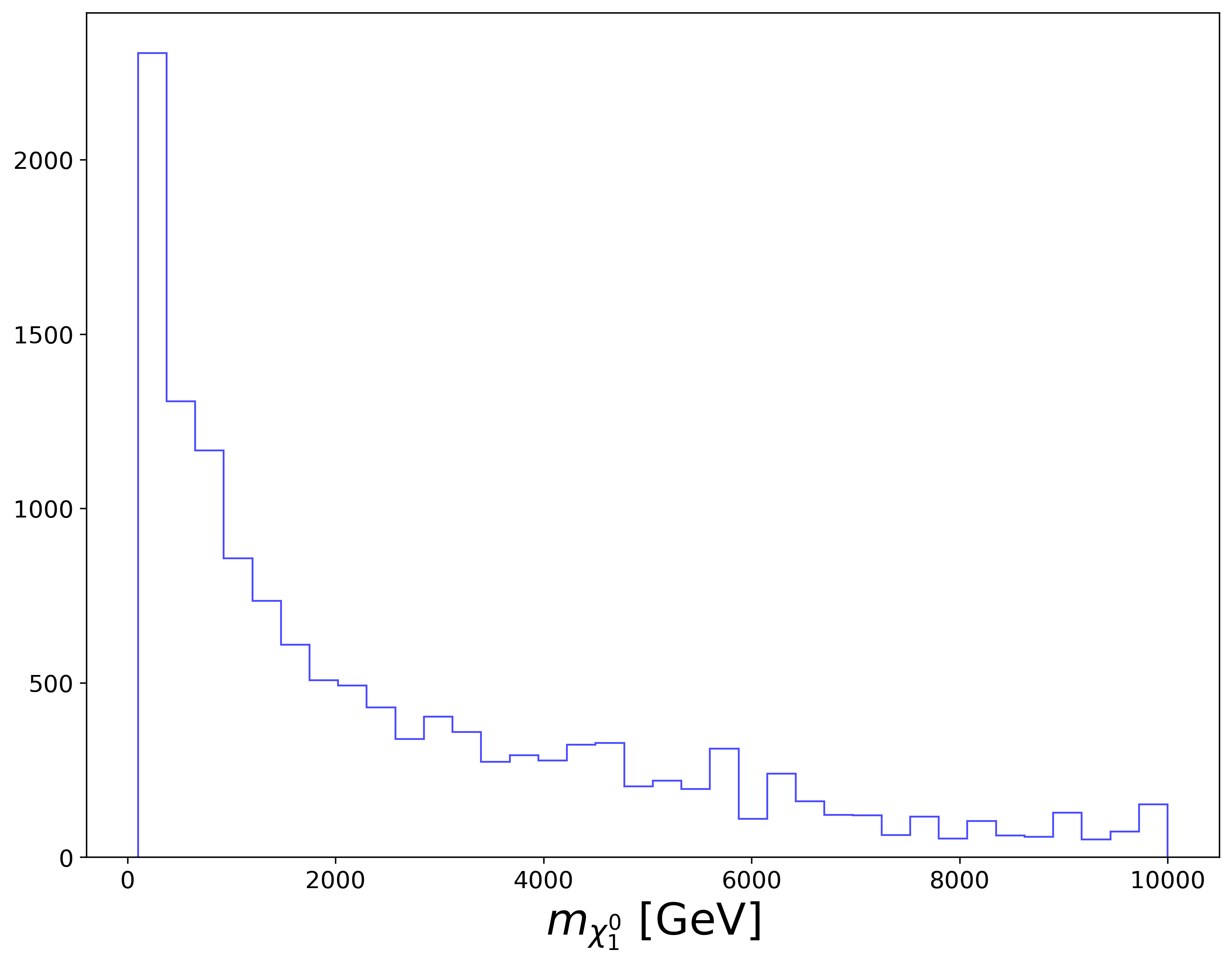}    
    ~~~~~~\includegraphics[width=0.46\textwidth]{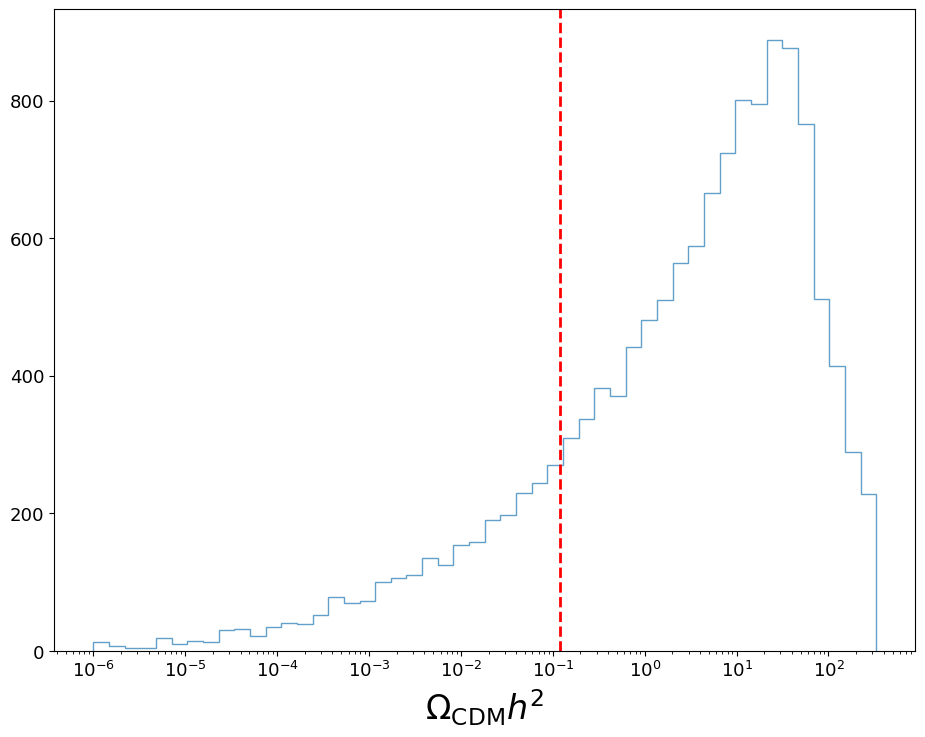}
    \caption{Mass distribution of the fermion FIMP $\chi_1^0$ (left) and distribution of the associated relic density $\Omega_{\rm CDM}h^2$ in the considered model obtained from a random scan yielding 13\,544 parameter points within the parameter space defined in Table \ref{Tab:DM_parameter_range}. The red line indicates the central value $\Omega_{\rm CDM}h^2 = 0.12$ of Eq.\ \eqref{Eq:Planck}.} 
    \label{Fig:Fermion_DM:Hist_omegah2}
\end{figure}

Coming to the case of fermion singlet dark matter, our random scan has given a total of 13\,544 accepted parameter points. The associated distributions of the dark matter mass and the relic density are shown in Fig.\ \ref{Fig:Fermion_DM:Hist_omegah2}. As for the scalar case, the mass distribution peaks towards the lower end of the scanned interval. However, in contrast to the scalar case, the relic density peaks at relatively high values, and only relatively few points (about 19\%) are found below the upper limit of Eq.\ \eqref{Eq:Planck}. The FIMP production cross-section in this case is thus higher as compared to the scalar case. As discussed in the scalar case, there is no significant correlation between the FIMP mass and the relic density, nor between the latter and the reheating temperature.

\begin{figure}[tb]
    \centering
    \includegraphics[width=0.455\textwidth]{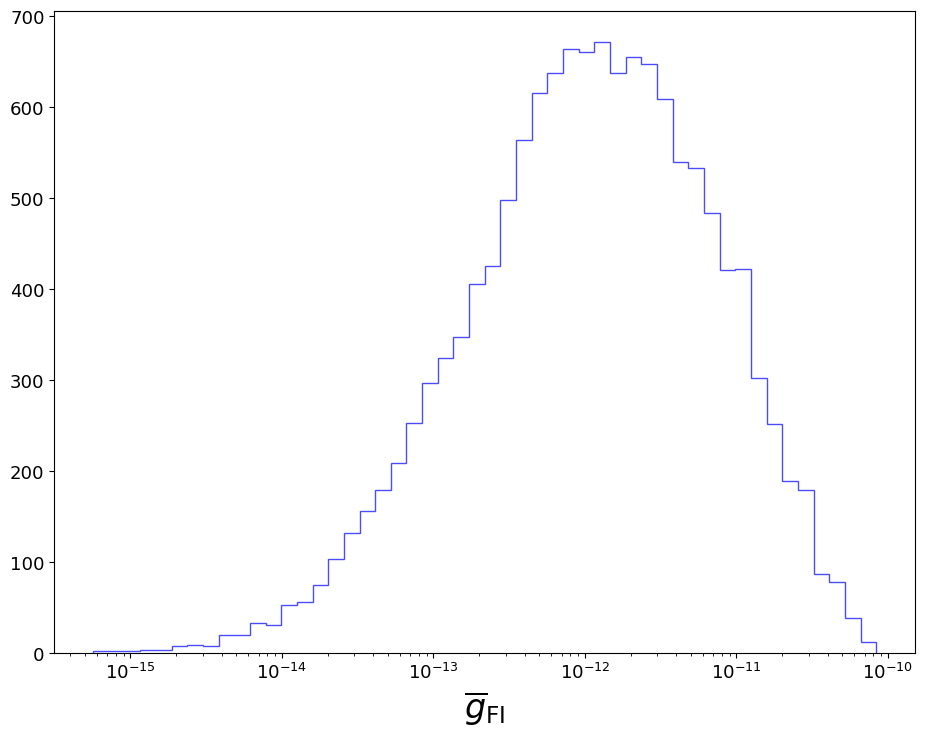}
    ~~~~
    \includegraphics[width=0.48\textwidth]{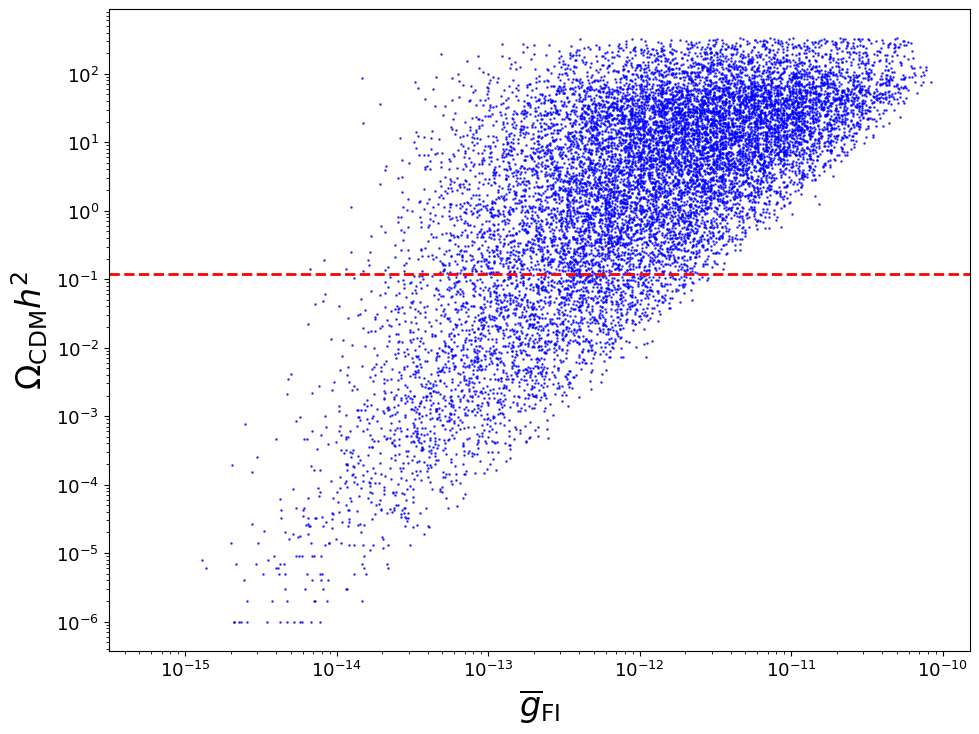}
    \caption{Left: Distribution of the geometric mean $\bar{g}_{\mathrm{FI}}$ of the couplings participating in the fermion FIMP production. Right: Correlation of the geometric mean of the scalar couplings $\bar{g}_{\mathrm{FI}}$ and the relic density $\Omega_{\rm CDM} h^2$ for the fermion FIMP $\chi^0_1$. Both plots are obtained from a random scan yielding 13\,544 viable parameter points featuring fermion FIMP DM.}
    \label{Fig:Fermion_DM:Scatter_geometrical_mean_gF_omegah2}
\end{figure}

This being said, the FIMP relic density is naturally correlated with the involved couplings. In Fig.\ \ref{Fig:Fermion_DM:Scatter_geometrical_mean_gF_omegah2}, we show the distribution of the geometric mean value,
\begin{equation}
    \bar{g}_{\mathrm{FI}} = \big( g_F^{e} g_F^{\mu} g_F^{\tau} \big)^{1/3} \,,
    \label{Eq:Geometric_mean_gF}
\end{equation}
and the correlation between this geometric mean value and the relic density. The mean value peaks around $\bar{g}_{\mathrm{FI}} \sim 10^{-12}$, which is the order of magnitude required for successful freeze-in. However, above this value, the relic density is typically too high and the FIMP overcloses the Universe. For $\Omega_{\rm CDM} \leq 0.12$, the coupling mean value typically lies in the interval $\bar{g}_{\mathrm{FI}} \in [10^{-15}, 10^{-12}]$.

\begin{figure}[tb]
\centering
    \includegraphics[width=0.45\textwidth]{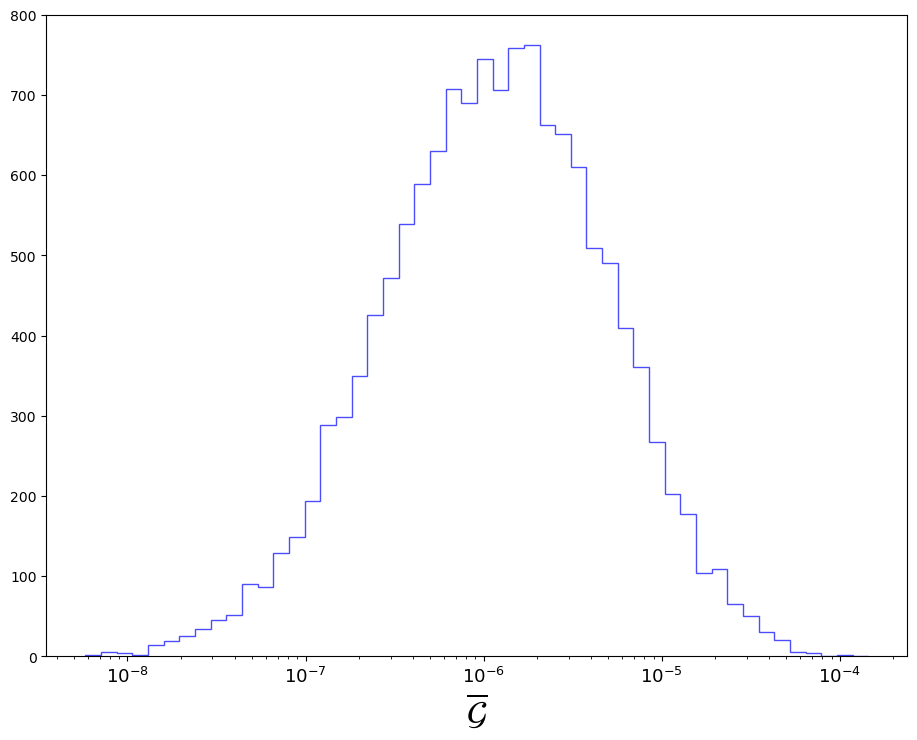}
    ~~~~
    \includegraphics[width=0.48\textwidth]{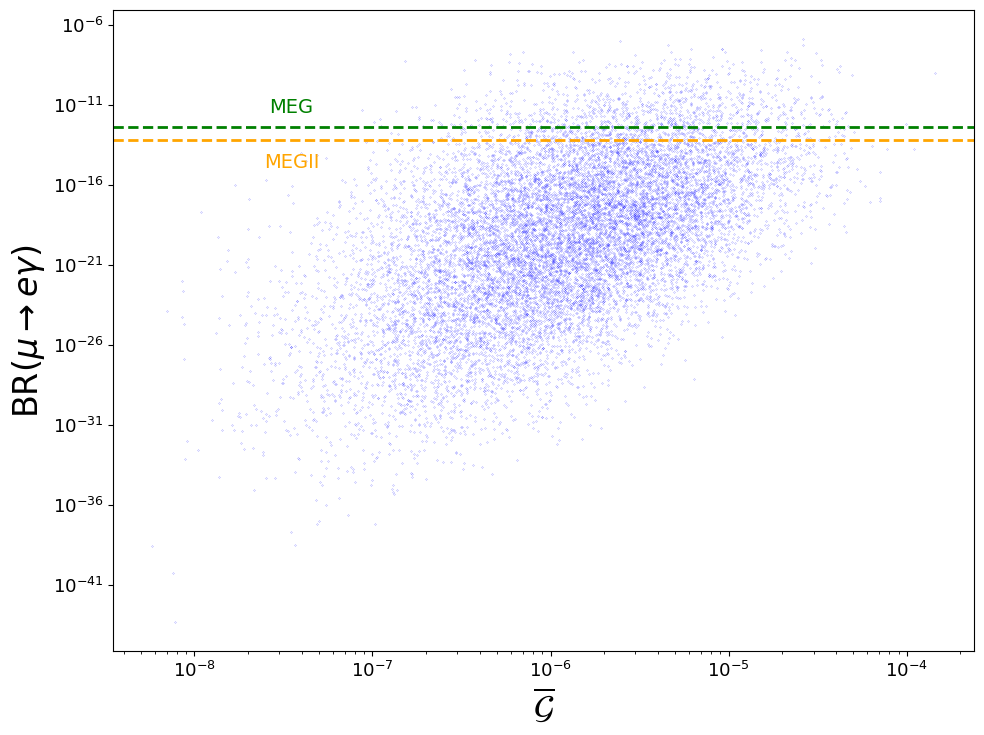}
    \caption{Left: Distribution of the modulus of the geometric mean of the couplings $\bar{\mathcal{G}}$ defined in Eq.\ \eqref{Eq:Modulus_geometric_mean_g}. Right: Correlation of the modulus of the geometric mean of the couplings $\bar{\mathcal{G}}$ and the branching ratio of the decay $\mu\to e\gamma$. Both plots are obtained from a random scan yielding 13\,544 viable parameter points featuring fermion FIMP DM.} 
    \label{Fig:Fermion_DM:Histo_modulus_g_geometric}
\end{figure}    

\begin{figure}[h!]
    \centering
    \includegraphics[width=0.47\textwidth]{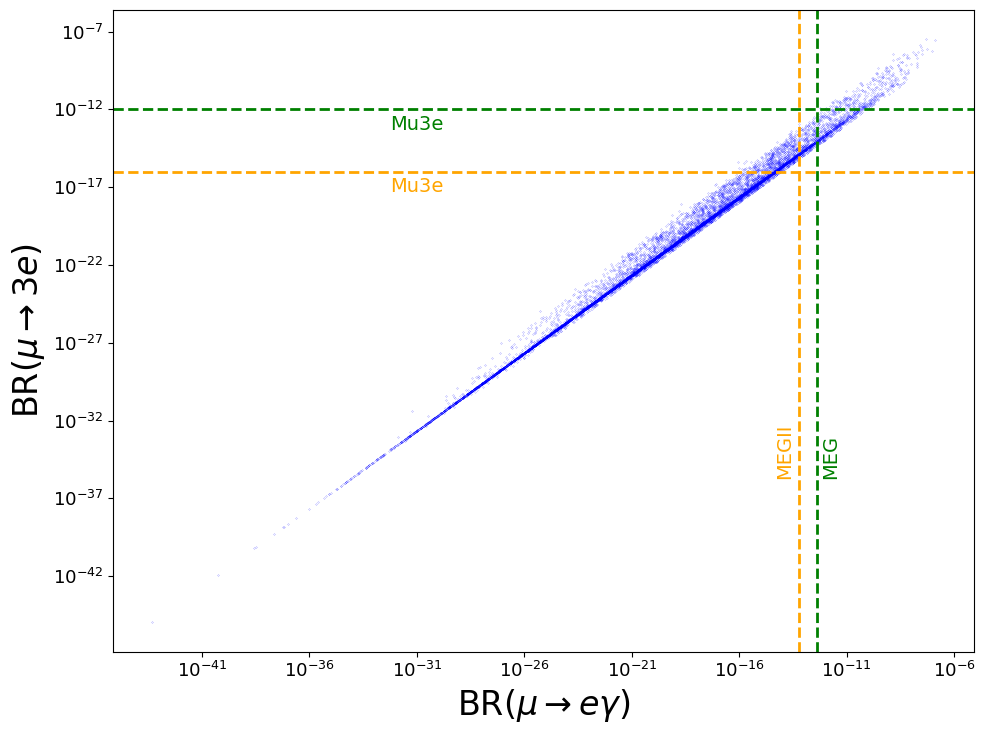}
    ~~~~\includegraphics[width=0.47\textwidth]{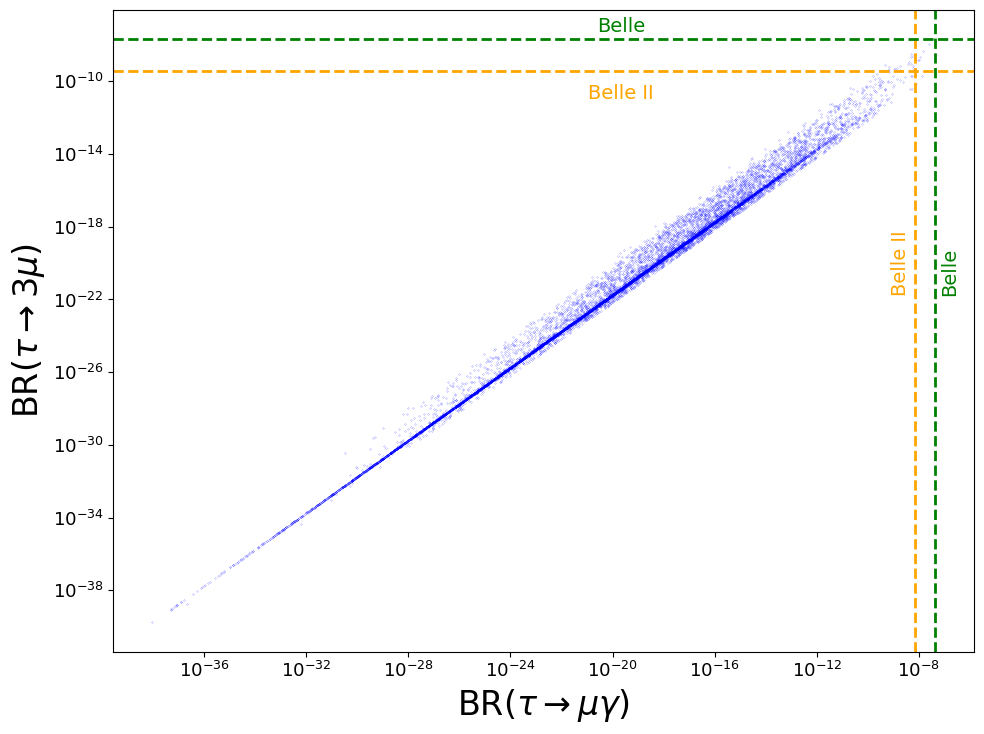}\\[2mm]
    \includegraphics[width=0.47\textwidth]{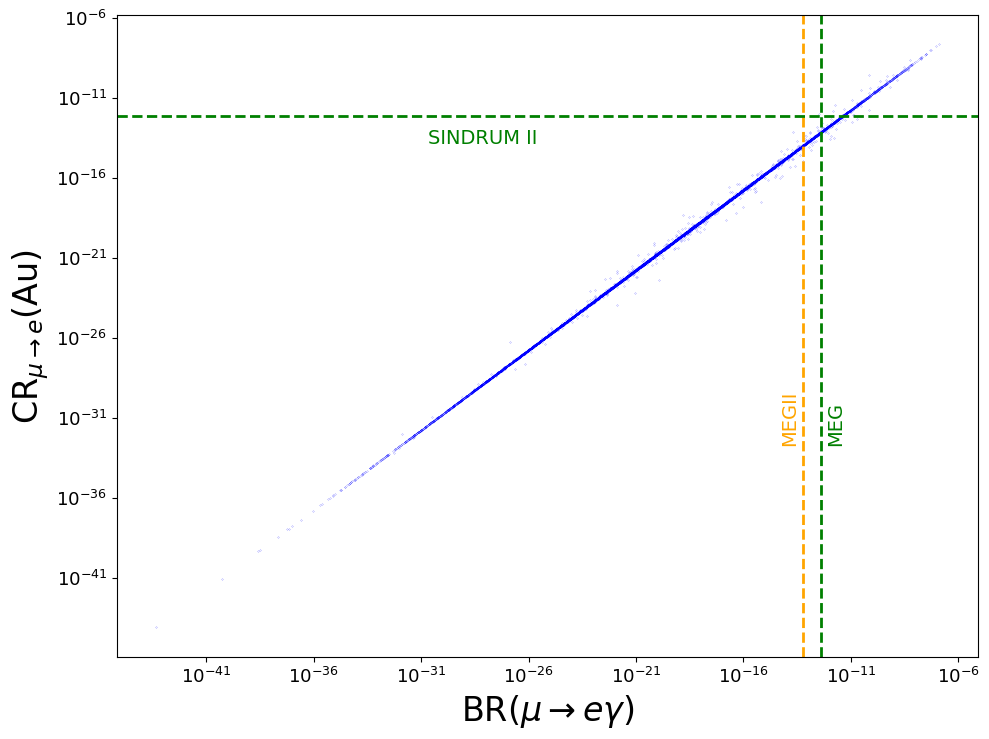}
    ~~~~\includegraphics[width=0.47\textwidth]{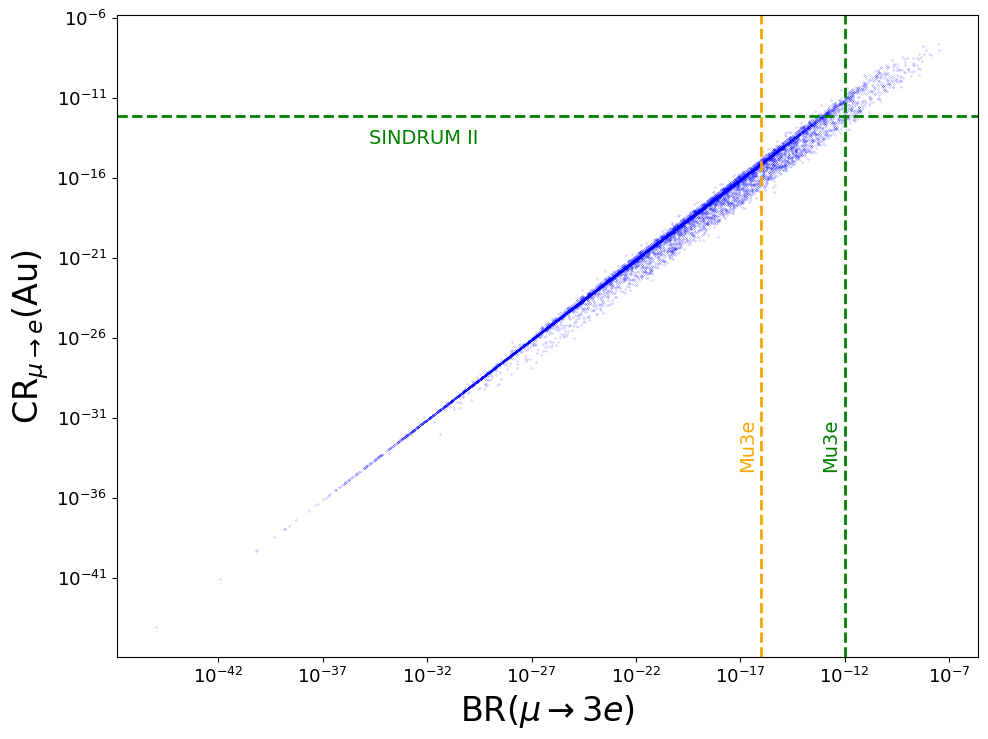}
    \caption{Upper left: Correlation of the respective branching ratios of the decays $\mu \to e \gamma$ and $\mu \to 3e$. Upper right: Same for the decays $\tau \to \mu \gamma$ and $\tau \to 3 \mu$. Lower left: Correlation between the $\mu-e$ CR in Au nuclei and BR of the decays $\mu \to e \gamma$. Lower right: Correlation between the $\mu-e$ CR in Au nuclei and BR of the decays $\mu \to  3e$. All plots are obtained from a random scan yielding 13\,544 viable parameter points featuring fermion FIMP DM. Green and orange dashed lines indicate current and future experimental limits, respectively.}
    \label{Fig:Fermion_DM:Scatter_BR_CR}
\end{figure}

Finally, we explore the phenomenology of lepton flavour violating observables in the present case of fermion FIMP DM. As stated in Sec.\ \ref{Subsec:Scalar_DM}, the BRs are only sensitive to the LH couplings $g^{\alpha}_{F}$ and $g^{\alpha j}_{\Sigma}$, $\alpha = e,\mu,\tau$ and $j = 1,2$ of the Lagrangian defined in Eq.\ \eqref{Eq:InteractionLagrangian}. The distribution of the geometric mean of these couplings, as defined in Eq.\ \eqref{Eq:Modulus_geometric_mean_g} is displayed in the left panel of Fig.\ \ref{Fig:Fermion_DM:Histo_modulus_g_geometric}. The mean value peaks around $10^{-6}$, which is about three orders of magnitude lower as in the scalar case discussed above. This is explained by the fact that, here, certain couplings are involved in LFV processes and in the FIMP production, namely the couplings $g_F^{\alpha}$ ($\alpha=e,\mu,\tau$). In the right panel of Fig.\ \ref{Fig:Fermion_DM:Histo_modulus_g_geometric}, we show $\bar{g}_{\mathrm{FI}}$ against the $\mu\to e\gamma$ branching ratio. As can be seen, no direct correlation is found, underlining the influence of the remaining parameters, such as the involved scalar and fermion masses. 

As in the scalar FIMP case, the correlations between the different LFV observables exhibit rather clean correlations, as can be seen in the examples shown in Fig.\ \ref{Fig:Fermion_DM:Scatter_BR_CR}. In this case, even more points satisfy the experimental limits (e.g., about 93\% for $\mu\to e\gamma$), which can be traced to the fact that, due to the involvement of some of them in the FIMP production, the couplings are overall slightly smaller, especially in cases of a lower relic density. Few points (about 3\% for $\mu\to e\gamma$) will be challenged by future experiments, while large parts of the available parameter space remain available with respect to lepton flavour violation. As for the scalar FIMP case, no points within the considered parameter regions are excluded by $\tau-\mu$ transitions.

%% file: tex_files/conclusion.tex
\section{Conclusion}
\label{Sec:Conclusion}

We have studied a scotogenic framework, where the Standard Model is extended by a scalar singlet, a scalar doublet, a fermion singlet, and a fermion triplet, assuming the freeze-in mechanism to explain today's dark matter (DM) abundance in the Universe. Allowing each singlet to be the candidate for a \dblquote{Feebly Interacting Massive Particle} (FIMP), we have scanned the model parameter space using the numerical codes {\tt SPheno} and {\tt micrOMEGAs}, the latter allowing to compute the DM relic density. 

Our scan shows that, in the considered model and within the considered parameter ranges, the relic density does not depend on the reheating temperature nor the FIMP mass. The relic density is rather sensitive to the couplings of the FIMP with the other particles in the model. In case of a scalar singlet FIMP, the relic density constraint given by Planck is satisfied in large regions of the considered parameter space. In the case of a fermion singlet FIMP, the relic density tends to be higher, and less points are in agreement with the Planck limit. 

Moreover, we have shown that the predictions for lepton flavour violating (LFV) processes, particularly $\mu-e$ transitions, do not exceed the experimental limits in a large portion of the parameter space. While a certain part will be challenged by future experimental data, the predictions remain below the projected limits for an important part of the model parameter space. 

While the present analysis is based on a pure random scan, in a next step the study could be taken to a more efficient way of exploring the parameter space, such as, e.g., a Markov Chain Monte Carlo scan (as in Refs.\ \cite{Sarazin:2021nwo, deNoyers:2024qjz}), or a Deep Neural Network approach (as in Ref.\ \cite{deNoyers:2025ije}). This would allow to extensively test the parameter space against a large number of constraints, including -- in addition to the previous study -- also dark matter direct detection, the Higgs mass measurement, neutrino masses, and all available LFV constraints. Based on such a scan, detailed comparisons between the freeze-in and the freeze-out pictures will be possible, e.g., on the level of physical masses and related observables at colliders. 

A further improvement would be the implementation of the Casas-Ibarra parametrization \cite{CasasIbarra2001, Herrero-Garcia:2025aox} of the neutrino sector, allowing to satisfy the constraints on the neutrino masses and mixing angles. While this can swiftly be done for the case of the scalar FIMP, it is more challenging for the fermion FIMP, as the concerned couplings contribute to both the neutrino masses and the FIMP production in this case. The necessary coupling hierarchy could be induced through methods similar to the one presented in Ref.\ \cite{Alvarez:2023dzz}, where the hierarchy is implemented on the level of the rotation matrix involved in the calculation of the couplings from the physical neutrino parameters.  

In summary, although there is room of improvement on the level of the performed parameter scan, our study shows that the FIMP scenario can easily be realized in typical scotogenic extensions of the Standard Model, provided that they include a singlet, which is a suitable candidate for FIMP DM. While the DM relic density constraint can be met by imposing relatively low coupling values, large parts of the parameter space are in agreement with current and future limits on lepton flavour violating processes. 